\documentstyle[psfig,conf_iap]{article}
\newcommand{\Lya}{Ly$\alpha$}

\begin{document}
\heading{
The Absorption Spectrum of the QSO PKS 2126-158 
($z_{\rm em} = 3.27$) and the Clustering Properties of Metal Systems
} 
\par\medskip\noindent
\author{%
Valentina D'Odorico$^1$
}
\address{%
International School for Advanced Studies, SISSA, via Beirut 
2-4, I-34014 Trieste, Italy
}
\begin{abstract}
We report the detection of a significant clustering signal 
on scales $\Delta v \lsim 300$ km s$^{-1}$, 
for a sample of 71 C~IV absorbers.
This result confirms previous findings and, together with data 
at lower column density, it points to a trend of decreasing 
clustering amplitude with decreasing column density, analogous 
to that observed for \Lya\ lines. 
\end{abstract}
\section{Observations and data analysis}

Spectra of the $z_{em} = 3.268$ quasar PKS 2126--158 have been obtained
at the ESO NTT telescope in the range $\lambda\lambda 4300-6620$ \AA\ with a 
resolution $R\simeq27000$ 
and an average $s/n$ ratio $\simeq 25$ per resolution element.
%
%
12 metal systems have been identified, two of which were previously
unknown. 
An enlarged sample of C~IV absorptions (71 doublets) has been used 
to analyze the statistical properties of this class 
of absorbers.
A full report on the data and an extended version of the 
discussion can be found in the paper \cite{Vale97}.

\section{Clustering of C~IV clouds}
We have adopted as a statistical tool the two-point correlation 
function (TPCF) computed in the velocity space.
The reference Poissonian distribution of pairs is obtained 
by averaging 5000 numerical simulations of the observed number of 
redshifts, randomly generated in the same 
redshift range as the data, according to the cosmological distribution 
$\propto (1+z)^{\gamma}$ where $\gamma=-1.2$ 
(\cite{Sarg88}). 

\begin{figure}
\centerline{\vbox{
\psfig{figure=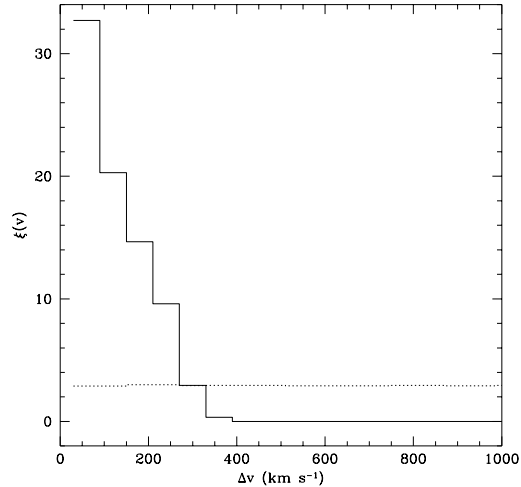,height=7.cm}
}}
\caption[]{\label{fig:corr2} TPCF at 
small velocity separations 
for the C~IV systems (binned at 60 km s$^{-1}$). 
The dotted line shows the 95\% confidence level. 
}
\end{figure}

The TPCF for the C~IV clouds, in the range $30 \le \Delta v \le 10000$ 
shows a strong clustering signal at small velocity separations 
(Fig.~\ref{fig:corr2}) 
while, at larger scales no significant signal is found. 
In the velocity interval $30 \le \Delta v \le 500$ km s$^{-1}$, 
66 pairs are observed, while $\sim 6$ are predicted for 
a homogeneous distribution, 
$\xi (30< \Delta v<500\ {\rm km\ s}^{-1}) = 10\pm 1$.
Similar results, with a significant correlation on scales up to 
$\sim 300 $ km s$^{-1}$, have been obtained in the works by \cite{PB94}, 
\cite{Wom96} and \cite{SC96}, 
carried out at comparable resolution. 
Degrading the resolution to that of previous works based on 
larger samples (\cite{Sarg88}; \cite{Heisler89}) 
a compatible result is obtained.


\section{Correlations with the column density}

Correlations between total equivalent widths and number of components 
and between total equivalent widths and velocity spread of the individual 
C~IV systems are observed. 
The TPCF for lower column density C~IV absorption systems 
shows a weaker signal than the present data 
(\cite{Wom96}; \cite{SC96}).
These three pieces of evidence suggest a trend of decreasing clustering 
amplitude with decreasing column density. An analogous behaviour 
has been observed for \Lya\ lines by \cite{Cris97}.  

\acknowledgements{I would like to thank S. Cristiani, S. D'Odorico, 
E. Giallongo and A. Fontana for their support and useful discussions.}


%
A
\vfill
\end{document}